\documentclass[aps,prx,reprint, amsmath, amssymb,superscriptaddress]{revtex4-1}

\pdfoutput=1
\usepackage[utf8]{inputenc}
\usepackage{bm}
\usepackage[english]{babel}
\usepackage[T1]{fontenc}
\usepackage{mathrsfs}
\usepackage[retainorgcmds]{IEEEtrantools}
\usepackage{amsmath}
\usepackage{amssymb}
\usepackage{color}
\usepackage{amsfonts}
\usepackage{times,txfonts}
\usepackage{nicefrac}
\usepackage[colorlinks=true,linkcolor=blue,urlcolor=blue,citecolor=blue,pdfusetitle]{hyperref}

\usepackage{float}
\usepackage{epsfig} 
\usepackage{subfigure}
\graphicspath{{Images/}}

\usepackage{physics}
\usepackage{braket}
\usepackage{amssymb}
\usepackage[amssymb]{SIunits}

\usepackage{xcolor}

\begin{document}

\title{%Dynamical stabilisation of multipartite 
Entanglement enhanced metrology with quantum many-body scars}
%\title{Quantum many-body scars: not so useless after all}

\author{Shane Dooley}
\email{shdooley@tcd.ie}
\affiliation{Department of Physics, Trinity College Dublin, Dublin 2, Ireland}
\author{Silvia Pappalardi}
\email{silvia.pappalardi@ens.phys.fr}
\affiliation{Laboratoire de Physique de l’\'Ecole Normale Sup\'erieure, ENS, Universit\'e PSL, CNRS,  Sorbonne Universit\'e, Universit\'e de Paris, F-75005 Paris, France}
\author{John Goold}
\email{gooldj@tcd.ie}
\affiliation{Department of Physics, Trinity College Dublin, Dublin 2, Ireland}

\begin{abstract}
 %Local observables of nonintegrable quantum many-body systems generally thermalise under time evolution from simple initial conditions. 
 Although entanglement is a key resource for quantum-enhanced metrology, not all entanglement is useful. For example in the process of many-body thermalisation, bipartite entanglement grows rapidly, naturally saturating to a volume law. This type of entanglement generation is ubiquitous in nature but has no known application in most quantum technologies. The generation, stabilisation and exploitation of genuine multipartite entanglement, on the other hand, is far more elusive yet highly desirable for metrological applications. Recently it has been shown that quantum many-body scars can have extensive multipartite entanglement. However the accessibility of this structure for real application has been so far unclear. In this work, we show how systems containing quantum many-body scars can be used to dynamically generate stable multipartite entanglement, and describe how to exploit this structure for phase estimation with a precision that beats the standard quantum limit. Key to this is a physically motivated modification of a Hamiltonian that generates a variety of multipartite entangled states through the dynamics in the scar subspace. 
\end{abstract}

\maketitle

\textbf{Introduction-} When a simple initial condition is evolved with a generic locally interacting many-body Hamiltonian, local observables are expected to thermalize~\cite{DAl-16, Mor-18}. As a consequence of the evolution, entanglement entropy grows rapidly, saturating at a volume law \cite{Cal-05, Chi-06, Kim-13} consistent with the thermal stationary value. This type of entanglement scaling, despite being ubiquitous \cite{Pag-93}, is known not to be useful in many quantum technologies \cite{Gro-09}.
On the other hand, multipartite entanglement -- witnessed by the quantum Fisher information (QFI) \cite{Tot-12, Hyl-12, Pez-09} -- is well known to be a crucial resource for quantum enhanced metrology \cite{Bra-94, Gio-11, Par-11, Pet-11, Pez-18}. 
In fact, the use of highly multipartite entangled states allows one to overcome classical limitations in quantum phase-estimation protocols \cite{Hol-03, Gio-11}. However, generally speaking, the fundamental challenge in creating and maintaining such states is that the dynamics of locally interacting many-body systems are thermalising, which leads to states that are ``too entangled to be useful'' \cite{Gro-09}.

The usual approach to solving the problem of thermalisation is to isolate every particle as much as possible, suppressing all unwanted interactions with other particles and with the wider environment. This approach has drawbacks: it can be difficult for large systems, and often requires the particles to be spatially well separated, limiting the overall size of the multipartite entangled state. Recent experimental \cite{Ber-17, Su-22, Zha-22} and theoretical work \cite{Tur-18, Ser-21, Mou-22} has uncovered a new mechanism -- weak ergodicity breaking -- that can prevent thermalisation in nonintegrable many-body systems. At the heart of this mechanism are quantum many-body scars (QMBS) -- rare, nonthermal eigenstates in a spectrum of otherwise thermal eigenstates. The theoretical properties of these eigenstates is currently an active area of research \cite{Shi-17, Tur-18, Doo-20b, Ser-21, Pap-21, Reg-21, Doo-22a, Sch-22, Cha-22, Mou-22}. For instance, although they display sub-volume bipartite entanglement it was recently shown that QMBS can have a non-trivial multipartite entanglement structure, embodied by an extensive scaling of the quantum Fisher information density~\cite{Des-21}. This raises the tantalising possibility of exploiting QMBS to dynamically generate multipartite entangled states, even in strongly interacting many-body systems. It is not practical to prepare a high energy eigenstate of a strongly interacting many-body system in the laboratory so at first glance it would seem difficult to try to exploit this entanglement structure for quantum enhanced metrology. Indeed recently it has been shown that QMBS can be exploited for robust quantum sensing, even in the presence of strong interactions \cite{Doo-21a} and QMBS have been discussed in connection to spin squeezing in Refs. \cite{Win-22, Com-22} but no protocol has been devised which directly exploits the entanglement structure. 

% In fact not only the Fisher information density experimentally accessible it also raises the tantalising possibility of exploiting QMBS to generate multipartite entangled states, even in strongly interacting many-body systems

The main result of this work is to offer a prescription for robust multipartite entanglement generation through time evolution with Hamiltonians that posses QMBS. Furthermore, we demonstrate how this structure can be exploited for quantum enhanced metrology. The key is a Hamiltonian modification that dynamically generates multipartite entanglement in the scar subspace, at the same time exploiting the QMBS to overcome the notorious fragility of multipartite entanglement to local interactions. Illustrating the idea with an example, we show that much larger multipartite entangled states can be generated in models with QMBS, in comparison to similar models without QMBS. To show the potential for quantum technologies, we propose a metrological scheme for phase estimation. Again, we see significantly better estimation precision in a model with QMBS, compared to a similar model without QMBS. \\

% We overcome the notorious fragility of multipartite entanglement and difficulty in stabilizing it, by suggesting a modified Hamiltonian which dynamically generates it in the scar subspace. 

\textbf{Exact many-body scars and multipartite entanglement-} Mark, Lin and Motrunich \cite{Mar-20} proposed a general framework for models with QMBS, called the spectrum-generating algebra (SGA) framework. The main ingredients are a linear subspace $\mathcal{S} \subset \mathcal{H}$ of the Hilbert space and an operator $\hat{Q}^+$ that obeys the following properties: (i) $\hat{Q}^+$ preserves the subspace $\mathcal{S}$, i.e., $(\hat{\mathbb{I}} - \hat{\mathcal{P}}_\mathcal{S}) \hat{Q}^+ \hat{\mathcal{P}}_\mathcal{S} = \hat{\mathcal{P}}_\mathcal{S} \hat{Q}^+ (\hat{\mathbb{I}} - \hat{\mathcal{P}}_\mathcal{S}) = 0$, where $\hat{\mathcal{P}}_\mathcal{S}$ is the projector into $\mathcal{S}$, and (ii) $\hat{Q}^+$ is a raising operator in $\mathcal{S}$, i.e.,  \begin{equation} 
\hat{\mathcal{P}}_{\mathcal{S}} \left( [\hat{H}, \hat{Q}^+] - \omega \hat{Q}^+ \right) \hat{\mathcal{P}}_{\mathcal{S}} = 0 . 
\label{eq:creation_in_S} 
\end{equation} 
If the subspace $\mathcal{S}$ contains a Hamiltonian eigenstate $\ket{\mathcal{S}_0} \in \mathcal{S}$, with $\hat{H}\ket{\mathcal{S}_0} = E_0\ket{\mathcal{S}_0}$, then the states $\ket{\mathcal{S}_j} = \mathcal{N}_j (\hat{Q}^+)^j \ket{\mathcal{S}_0} \in \mathcal{S}$ (where $\mathcal{N}_j$ is a normalisation factor) are also Hamiltonian eigenstates $\hat{H}\ket{\mathcal{S}_j} = (E_0 + j\omega)\ket{\mathcal{S}_j}$. When the Hamiltonian $\hat{H}$ is found to be nonintegrable (e.g., through analysis of the energy level spacing statistics \cite{Tur-18}) these states $| \mathcal{S}_j \rangle$ are regarded as quantum many-body scars.

It was recently shown that the scars have extensive multipartite entanglement, in contrast to the thermal eigenstates that make up the remainder of the spectrum of $\hat{H}$. The degree of multipartite entanglement of a quantum state $\hat{\rho}$ can be probed with the quantum Fisher information (QFI)~\cite{Hel-69, Tth-14, Pez-14}:
\begin{equation} \mathcal{F} (\hat{O}, \hat{\rho}) = 2 \sum_{n,m} \frac{(p_n - p_m)^2}{p_n + p_m} |\langle n | \hat{O} | m \rangle |^2 \leq 4 \text{Var}_{\hat{\rho}}\hat{O} , 
\end{equation} 
with respect to an appropriately chosen observable $\hat{O}$. Here, $p_n$, $\ket{n}$ are the eigenvalues and eigenstates of $\hat{\rho}$ and the upper bound $\text{Var}_{\hat{\rho}}\hat{O} = \text{Tr}(\hat{O}^2 \hat{\rho}) - [\text{Tr}(\hat{O} \hat{\rho})]^2$ is achieved for pure states $\hat{\rho} = \ket{\psi}\bra{\psi}$. For an $N$ particle system, and a collective observable $\hat{O} = \frac{1}{2}\sum_{n=1}^N \hat{o}_n$ (sum of local ones $\hat{o}_n$), a QFI density satisfying: \begin{equation} f (\hat{O}, \hat{\rho}) \equiv \frac{\mathcal{F} (\hat{O}, \hat{\rho})}{N} > m , \end{equation} indicates that at least $m+1$ particles of the system are entangled ~\cite{Hyl-12, Tot-12, Pez-09}. Importantly, the QFI is also a key quantifier of the usefulness of the state $\hat{\rho}$ in quantum metrology~\cite{Bra-94, Pet-11, Pez-18}: the precision in estimating a parameter $\varepsilon$, encoded in the state $\hat{\rho}_\varepsilon = e^{-i\varepsilon\hat{O}}\hat{\rho}e^{i\varepsilon\hat{O}}$, is bounded by the Cram\'{e}r-Rao inequality $\delta\varepsilon \geq 1/\sqrt{\nu\mathcal{F}}$, where $\nu$ is the number of independent repetitions of the measurement. 
Hence, while phase sensitivity with separable state is bounded by the  \emph{standard quantum limit} $\delta {\varepsilon}\leq \delta \varepsilon_{\text{SQL}} = 1/\sqrt{\nu N}$, the presence of multipartite entanglement results in an enhancement up to $\delta \varepsilon\leq \delta \varepsilon_{\text{HL}} = 1/\sqrt{\nu}N$, known as the \emph{Heisenberg limit} ~\cite{Hol-03, Gio-11}.
Recently, the QFI has been studied theoretically and experimentally in many-body systems due to its relation to thermal susceptibilities~\cite{Hau-16, Pap-17, Bre-20, Lau-21, Sch-21, DeA-21}. However in the case of locally interacting Hamiltonians, it is generally very challenging to generate quantum states with the Heisenberg scaling. 

\begin{figure}
    \begin{center}
	\includegraphics[width=\columnwidth]{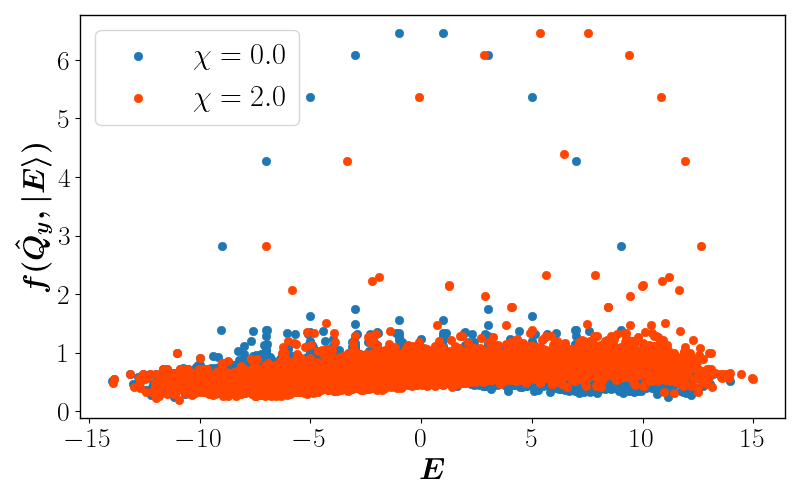}
		% Create a subtitle for the figure.
	\caption{The QFI density for the eigenstates $|E\rangle$ of the Hamiltonian $\hat{H}$ in Eqs. \ref{eq:H_0}, \ref{eq:H_int} (in its zero-momentum symmetry sector) and for the local observable $\hat{Q}^y = \frac{i}{2}(\hat{Q}^- - \hat{Q}^+)$. The QMBS have much larger QFI density than the thermal eigenstates. For $\chi = 0$ (blue) the QMBS are equally spaced in energy, but $\chi \neq 0$ (red) breaks the harmonic energy spacing. [Parameters: $N=11$, $\omega = 2$, $\lambda = 1$, $\gamma = 2$, $\eta=\pi/2$, $L=10$, $d=1$. For clarity in the plotted data, a small Hamiltonian perturbation $\hat{H}_\text{pert} = 10^{-5}\hat{Q}^x$ is also included here to break some degeneracies due to magnetization symmetry of $\hat{H}$.]}
	% 
	
		% Define the label of the figure. It's good to use 'fig:title', so you know that the label belongs to a figure.
	\label{fig:extensive_scar_entanglement}
	\end{center}
\end{figure}

Ref. \cite{Des-21} showed that any QMBS with nonzero energy-density has a QFI density with exactly this property:  extensive $f \sim N$, while for thermal eigenstates $f \sim 1$. Despite the extensive multipartite entanglement of the QMBS, dynamics through the scar subpsace are often though states with low entanglement. This is because, typically, the system is initialised in an easily prepared separable (or low-entanglement) state, and dynamics through the scar subspace does not significantly change the entanglement structure. In order to see this, we note that Eq.~\ref{eq:creation_in_S} above implies that the Heisenberg equation-of-motion for the operator $\hat{Q}^+_\mathcal{S} \equiv \hat{\mathcal{P}}_\mathcal{S} \hat{Q}^+ \hat{\mathcal{P}}_\mathcal{S}$ is linear inside the scarred subspace, $\frac{d}{dt} \hat{Q}^+_\mathcal{S} (t) = i\omega \hat{Q}^+_\mathcal{S} (t) \implies \hat{Q}^+_\mathcal{S} (t) = \hat{Q}^+_\mathcal{S}(0) e^{i\omega t}$. Assume that the initial state is a generalised coherent states $\ket{\theta,\phi} \equiv \hat{D}(\theta,\phi)\ket{\mathcal{S}_0}$, where $\hat{D}(\theta,\phi) = \exp\{\frac{\theta}{2}(\hat{Q}^+ e^{-i\phi} - \hat{Q}^- e^{i\phi}) \}$ is the unitary ``displacement'' operator and $\hat{Q}^- = (\hat{Q}^+)^\dagger$. Such a state is a superposition of QMBS, without any overlap with the other thermal eigenstates. Then, due to the linearity of the equation-of-motion for $\hat{Q}^+_\mathcal{S}$, the time-evolved state is: 
\begin{eqnarray} \ket{\psi(t)} &=& e^{-it\hat{H}}\ket{\theta,\phi}% \\ &=& 
%= \exp\left\{\frac{\theta}{2}[\hat{Q}^+ (t)e^{-i\phi} - \xi^* \hat{Q}^- (t)e^{i\phi} ]\right\} e^{-itE_0} \ket{\mathcal{S}_0} %\\ &=& 
= e^{-itE_0} \ket{\theta, \phi - \omega t} , \end{eqnarray} 
i.e., the time-evolved state is always a generalised coherent state in the QMBS subspace. If the coherent state displacement operator $\hat{D}(\theta,\phi)$ can be expressed as a product of local unitaries (as is often the case), then every coherent state $\ket{\theta,\phi} = \hat{D}(\theta,\phi)\ket{\mathcal{S}_0}$ has the same entanglement structure as $\ket{\mathcal{S}_0}$. In particular, if $\ket{\mathcal{S}_0}$ has low entanglement then the time-evolving state $\ket{\psi(t)}$ will too. We also note that the evolving state undergoes periodic revivals to the initial state with a period $T = 2\pi/\omega$, which is characteristic of most examples of dynamics in scarred many-body systems up to now. 

In order to generate multipartite entangled states in the scarred subspace $\mathcal{S}$ we consider adding a term 
\begin{equation}
    \label{eq:Hnl}
    \hat{H}_\text{nl} = \frac{\chi}{N} \hat{Q}^+ \hat{Q}^-
\end{equation} to a Hamiltonian that possesses a scarred subspace. Here $N$ is the number of particles. This additional term preserves the QMBS subspace, yet if $[\hat{Q}^+ \hat{Q}^-, \hat{Q}^+] \not \propto c_+ \hat{Q}^+ + c_- \hat{Q}^-$ (for complex coefficients $c_\pm$) then it also generates non-linear evolution. Moreover, as discussed in Appendix \ref{app:deriving_nonlinear_H}, a Hamiltonian term of the form $\hat{H}_\text{nl}$ can emerge naturally in the scar subspace through a linear coupling of the system to highly detuned ancillary system. To illustrate these ideas, we now provide a concrete example of a Hamiltonian with QMBS within the SGA framework.\\

\textbf{Model-} Consider a system of spin-1 particles on a $d$-dimensional cubic lattice with the Hamiltonian ${\hat{H}(\eta) = \hat{H}_0 + \hat{H}_\text{int}(\eta)}$, where: 
\begin{eqnarray} \hat{H}_0 &=& \frac{\omega}{2} \sum_{\vec{n}}\hat{S}_{\vec{n}}^z , \label{eq:H_0} \\ \hat{H}_\text{int}(\eta) &=& \sum_{\vec{n},\vec{n}'} \lambda_{\vec{n},\vec{n}'} ( e^{i\eta} \hat{S}_{\vec{n}}^+ \hat{S}_{\vec{n}'}^- + e^{-i\eta} \hat{S}_{\vec{n}}^- \hat{S}_{\vec{n}'}^+ ) . \label{eq:H_int} % + \frac{\Omega}{2} \sum_{\vec{n}} [(\hat{S}_{\vec{n}}^-)^2 + \text{h.c.}]
\end{eqnarray} 
Here $\hat{S}_{\vec{n}}^\pm$ are the spin-1 raising and lowering operators for the particle at the lattice site labelled by $\vec{n} \in \mathbb{Z}^d$, and $[\hat{S}_{\vec{n}}^+, \hat{S}_{\vec{n}'}^-] = \delta_{\vec{n},\vec{n}'} \hat{S}_{\vec{n}}^z$. For convenience, we assume that this interaction term has a power law decay $\lambda_{\vec{n},\vec{n}'} = \lambda / (a |\vec{n} -\vec{n}'|)^\gamma$, where $a |\vec{n} -\vec{n}'|$ is the distance between the particles at sites $\vec{n}$ and $\vec{n}'$ on the lattice, $a$ is the lattice spacing, and  $\gamma$ is the range of the power law decay. If we assume that the lattice is contained in a fixed volume $V = L^d$ then the total number of particles $N = (L/a + 1)^d \stackrel{L \gg a}{\approx} V / a^d$ can only be increased by decreasing the lattice spacing $a$ which, however, also results in an increased interaction strength $\lambda_{\vec{n},\vec{n}'}$. %\textcolor{blue}{Add the limit in which this becomes the XY + cit?}
An analysis of the energy level spacing statistics in a symmetry resolved subspace of $\hat{H}(\eta)$ shows that the Hamiltonian is quantum chaotic for all values of the phase $\eta$ \cite{Doo-21a}. However, for $\eta = \pm \pi/2$ the term $\hat{H}_\text{int}$ is a Dzyaloshinskii–Moriya interaction (DMI) and the total Hamiltonian has a set of $N+1$ quantum many-body scars $\ket{\mathcal{S}_j} = \mathcal{N}_j (\hat{Q}^+)^j \ket{\mathcal{S}_0}$ 
with
\begin{equation}
    \label{eq:raisingModel}
    \hat{Q}^+ = \sum_{\vec{n}} (\hat{S}_{\vec{n}}^+)^2 \ ,\quad 
    \ket{\mathcal{S}_0} = \bigotimes_{\vec{n}}\ket{S_{\vec{n}}^z = -1} \ .
\end{equation}
%built on top of the fully polarized spin state $\ket{\mathcal{S}_0} = \bigotimes_{\vec{n}}\ket{S_{\vec{n}}^z = -1}$ with the raising operator $\hat{Q}^+ = \sum_{\vec{n}} (\hat{S}_{\vec{n}}^+)^2$. 
In this case, $\hat{H}(\eta=\pm \pi/2)$ fits within the SGA framework described above [cf. Eq.\eqref{eq:creation_in_S}], and the QMBS at nonzero energy density have an extensive multipartite entanglement (see blue markers in Fig. \ref{fig:extensive_scar_entanglement}). We note that the scarred Hamiltonian $\hat{H}(\eta=\pm \pi/2)$ is closely related to another spin-1 model that is known to host QMBS: the spin-1 XY magnet %introduced by Schecter and Iadecola 
\cite{Sch-19}\footnote{The two models are related by a unitary transformation if the graph with edges given by non-zero interactions $\lambda_{\vec{n},\vec{n}}$ is bipartite \cite{Mar-20b}}.

\begin{figure}
    \begin{center}
	\includegraphics[width=\columnwidth]{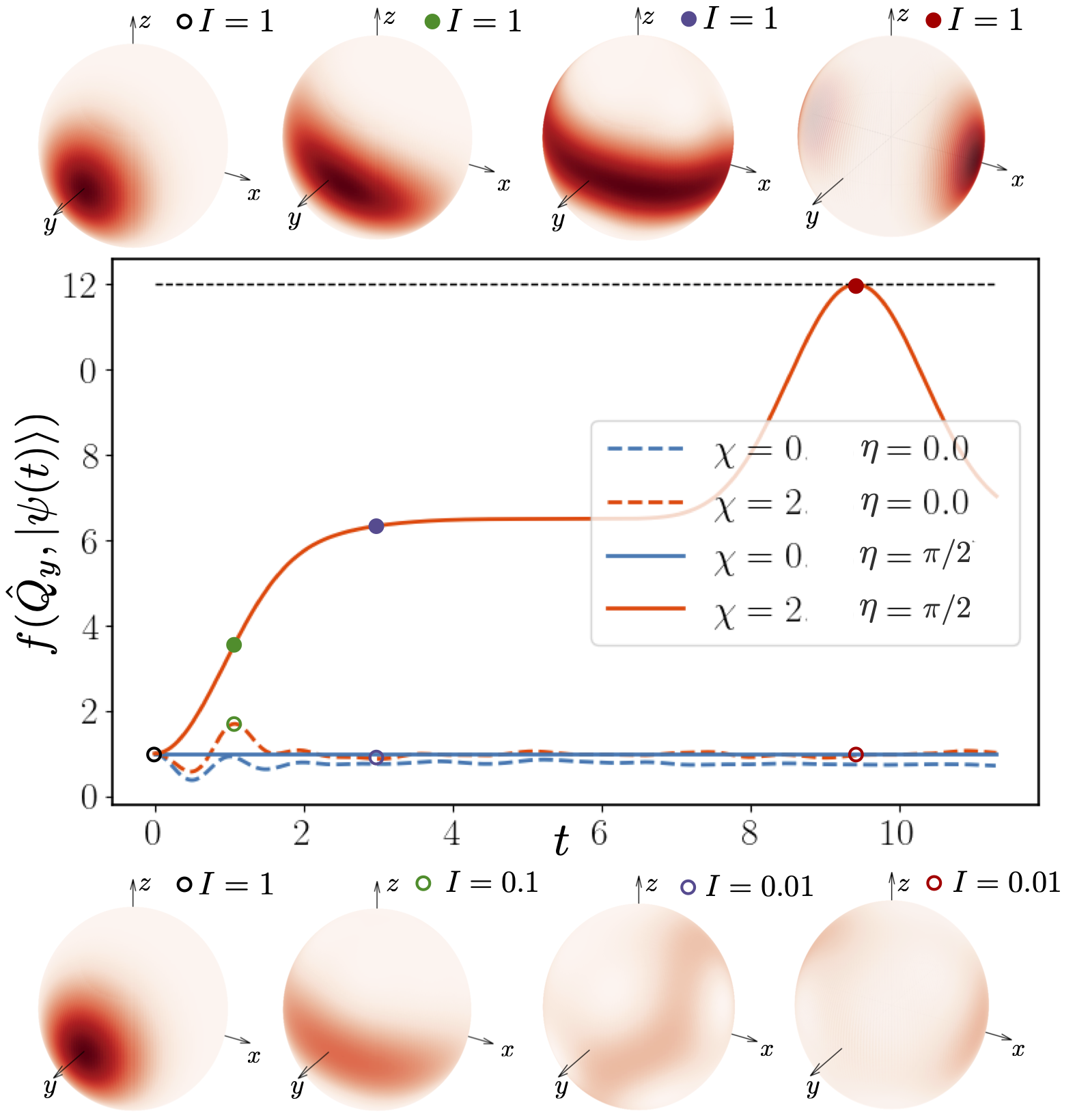}
		% Create a subtitle for the figure.
	\caption{The evolution of the QFI density by the Hamiltonian $\hat{H}$, with the initial product state $\ket{\psi(0)} = \ket{\theta = \frac{\pi}{2},\phi=0} = \bigotimes_{\vec{n}} [(\ket{+1_{\vec{n}}} + \ket{-1_{\vec{n}}})/\sqrt{2}]$. Without QMBS ($\eta = 0$) the dynamics never generates states with large $f$. For the Hamiltonian with QMBS ($\eta=\pi/2$) the QFI density remains small for $\chi = 0$, but $\chi \neq 0$ generates a variety of highly multipartite entangled states. The dashed black line indicates the Heisenberg limit, $f=N$. [Parameters: $N=12$, $\omega = 0$, $\lambda = 1$, $\gamma = 2$, $L=10$, $d=1$.]}
		% Define the label of the figure. It's good to use 'fig:title', so you know that the label belongs to a figure.
	\label{fig:qfi_dynamics}
	\end{center}
\end{figure}

It is straightforward to show that the operator $\hat{Q}^+$ given in Eq. \eqref{eq:raisingModel}, along with $\hat{Q}^-$ and $\hat{Q}^z = \frac{1}{2}[\hat{Q}^+, \hat{Q}^-]  = \frac{1}{2}\sum_{\vec{n}}\hat{S}_{\vec{n}}^z$ form an SU(2) algebra. The corresponding SU(2) spin coherent states $\ket{\theta,\phi} = \bigotimes_{\vec{n}}[\cos\frac{\theta}{2}\ket{-1_{\vec{n}}} + e^{-i\phi}\sin\frac{\theta}{2}\ket{+1_{\vec{n}}}]$ are separable product states of the spins. Preparing such a state initially and allowing it to evolve by $\hat{H}(\eta=\pm\pi/2)$ leads to periodic revivals and the system remains in a product state throughout the dynamics. This is reflected by the small, constant value of the QFI density as a function of time (see Fig. \ref{fig:qfi_dynamics}, solid blue line). So, despite the extensive multipartite entanglement of the QMBS, the dynamics involves only product states. For comparison, we also show the dynamics of the QFI density for evolution by the Hamiltonian $\hat{H}(\eta=0)$, i.e., for a similar model without QMBS. In that case the system approaches a thermalised state with $f \sim 1$ (Fig. \ref{fig:qfi_dynamics}, dashed blue line). \\

\textbf{Dynamical generation of entanglement-} 
%We now consider the addition of a term $\hat{H}_{\text{nl}} = \frac{\chi}{N} \hat{Q}^+ \hat{Q}^-$ to our model Hamiltonian, where $\hat{Q}^+ = \sum_{\vec{n}} (\hat{S}_{\vec{n}}^+)^2$ is the raising operator in the QMBS subspace. 
 We now consider the addition of the non-linear term Eq.\eqref{eq:Hnl} to our model $H(\eta)$. The total Hamiltonian thus reads
\begin{equation}
    \hat H_{\text{tot}}(\eta) = \hat H(\eta) + \hat{H}_{\text{nl}}  = \hat H(\eta) +\frac{\chi}{N} \hat{Q}^+ \hat{Q}^-
\end{equation}
where $\hat{Q}^+$ is the raising operator in the QMBS subspace defined in Eq. \eqref{eq:raisingModel}. 
Fig. \ref{fig:extensive_scar_entanglement} (red markers) shows the QFI density of the eigenstates of the scarred Hamiltonian $\hat H_{\text{tot}}(\eta=\pm \pi/2)$. %$\hat{H}(\phi=\pm\pi/2)+\hat{H}_\text{nl}$. 
We see that the addition of the nonlinear term preserves the existence of the QMBS, and their high QFI density, but destroys the harmonic spacing between them. Thus, upon dynamics from a spin coherent state it will generate a variety of multipartite entangled states.
Using the SU(2) commutation relations, $\hat{H}_\text{nl}$ can be rewritten as $\hat{H}_\text{nl} = \frac{\chi}{N}[\hat{Q}^2 - (\hat{Q}^z)^2 + \hat{Q}^z]$, where $\hat{Q}^2 = \frac{1}{2}\{ \hat{Q}^+, \hat{Q}^- \} + (\hat{Q}^z)^2$. 
%The one-axis twisting term $\sim (\hat{Q}^z)^2$ is known to generate a variety of multipartite entangled states through its dynamics, starting from a spin coherent state. 
The one-axis twisting term $\sim (\hat{Q}^z)^2$ is well known to dynamically generate multipartite entanglement in collective spin systems \cite{Kit-93, Aga-97, Chu-99, Doo-14}. 
Here, we show how this also holds in a many-body system with local interactions, provided that its Hamiltonian has QMBS. We initialize in a coherent state 
\begin{equation}
    \label{psi0}
    \ket{\psi_0} = \ket{\theta = \frac{\pi}{2}, \phi = 0} = \bigotimes_{\vec{n}} \frac{\ket{+1_{\vec{n}}} + \ket{-1_{\vec{n}}}}{\sqrt{2}}
\end{equation}
and consider its evolution with the total Hamiltonian $\hat H_{\text{tot}}(\eta)$. 
Fig. \ref{fig:qfi_dynamics} (red line) shows that the addition of the new term $\hat{H}_\text{nl} = \frac{\chi}{N} \hat Q^+ \hat Q^-$ dynamically generates highly multipartite entangled states. 
After an initial transient, the QFI of $\ket{\psi(t)}$ becomes compatible with that of a Dicke state whose QFI is super-extensive with $N$, in particular $f= N/2$.
We also notice that at large times $t^* \sim N/\chi$, quantum interference effects lead to the generation of macroscopic cat states, also known as Greenberger–Horne–Zeilinger (GHZ) state, for which the Heisenberg limit $f(t^*)=N$ is reached \cite{Aga-97, Chu-99, Doo-14, Pez-18}.  
For comparison, we also show the dynamics of the QFI density for the corresponding model without QMBS at $\eta = 0$ (dashed red line). In this case the QFI density is $f \sim 1$, indicating that the QMBS are crucial for generating multipartite entanglement in this model, starting from the same spin-coherent state.

% $t^*=\pi N/4$

% \textcolor{blue}{S: maybe we can chose between these two representations L $\xi$ or $\theta, \phi$}
This dynamical protocol is the key insight of the paper and it can be visualized on the generalized Bloch sphere indentified by the SU(2) algebra. States that overlap with the scars can be represented by the Husimi distribution $Q(\theta, \phi) = |\bra{\psi} \theta, \phi \rangle |^2$, where $|\theta, \phi \rangle $ is a coherent state on the sphere. The $Q$ is characterized by the property that its integral 
\begin{equation}
    \label{eq:integraHusi}
    I = \frac{N}{4 \pi} \int  d\theta d\phi \sin \theta\,\, Q(\theta, \phi) 
\end{equation}
is normalized ($I=1$) for states that live only on the Bloch sphere, while in general $0\leq I \leq 1$. Hence $I$ quantifies the localization of the state on the scarred subspace. In the absence of QMBS  (bottom panel of Fig.~\ref{fig:qfi_dynamics}), the initial coherent state spreads over all the Hilbert space and leaves the Bloch sphere, as shown by $I(t)<1$. On the other hand, when the state is evolved with $\hat H_{\text{tot}}(\eta=\pi/2)$ (upper panel of Fig.~\ref{fig:qfi_dynamics}), the $Q$ displays evolution with $I(t)=1$: it initially undergoes squeezing (green marker), it then spreads on the equator (purple marker) and it is eventually exhibits a GHZ state at $t^*$ (red marker).

\begin{figure}[t]
    \begin{center}
	\includegraphics[width=\columnwidth]{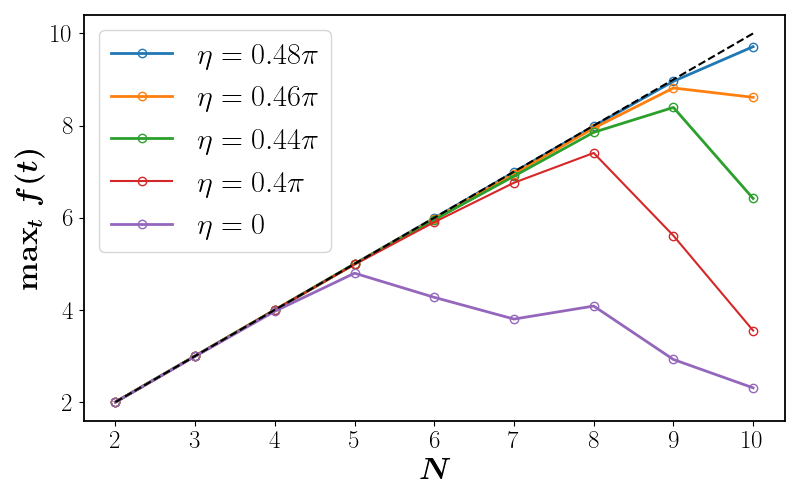}
		% Create a subtitle for the figure.
	\caption{The largest QFI density $\max_t f(t) = \max_t f(\hat{Q}_y, \ket{\psi(t)})$ that can be generated dynamically in a system of $N$ spins constrained to a fixed volume $V=L^d$. Since the volume is fixed, increasing the system size $N$ leads to higher density, stronger interactions, and inhibits the growth of the QFI density in the absence of QMBS. The dashed black line shows the Heisenberg limit $f=N$. [Parameters: $\omega = 1$, $\chi=2$, $\lambda = 1$, $\gamma = 2$, $d=1$, $\theta =\pi/2$, $\phi = 0$.]}
		% Define the label of the figure. It's good to use 'fig:title', so you know that the label belongs to a figure.
	\label{fig:max_qfi_vs_N}
	\end{center}
\end{figure}

%The dynamical generation of highly multipartite entangled states is not generic of many-body interacting systems, and it highly relies on the presence of the MBQS.
Another advantage of the generation of highly multipartite states using QMBS is shown by considering finite volume effects in realistic protocols. 
In our model Hamiltonian, we have assumed that the spins are on a square lattice with a fixed volume $V$. To increase the number of particles in the fixed volume therefore means increasing their density, which results in stronger interactions. Typically, this limits the size of multipartite entangled state that can be generated in the dynamics. This is shown in Fig. \ref{fig:max_qfi_vs_N}. We see that for small $N$ (corresponding to a low-density of particles) the QFI density that can be achieved actually approaches the Heisenberg limit $f=N$ for any value of $\eta$. However, in the absence of QMBS, larger values of $N$ (corresponding to an increased particle density), result in stronger interactions between the particles that inhibit the size of the multipartite entangled state. On the contrary, for a Hamiltonian with perfect QMBS ($\phi=\pm\eta/2$) the size of the multipartite entangled state can grow with the Heisenberg scaling $f \sim N$ despite the stronger interactions. \\

\textbf{Entanglement-enhanced quantum metrology-} The quantum Fisher information gives a bound to the estimation error that is only achievable for an optimal measurement. However, such optimal measurements are often impractical, particularly for large multipartite entangled states. Following Ref. \cite{Dav-16}, we propose that the multipartite entanglement generated in our model can be exploited for quantum metrology by a feasible ``echo'' measurement \cite{Mac-16, Hos-16, Lin-16, Nol-17, Hay-18}.
\begin{figure}
    \begin{center}
	\includegraphics[width=\columnwidth]{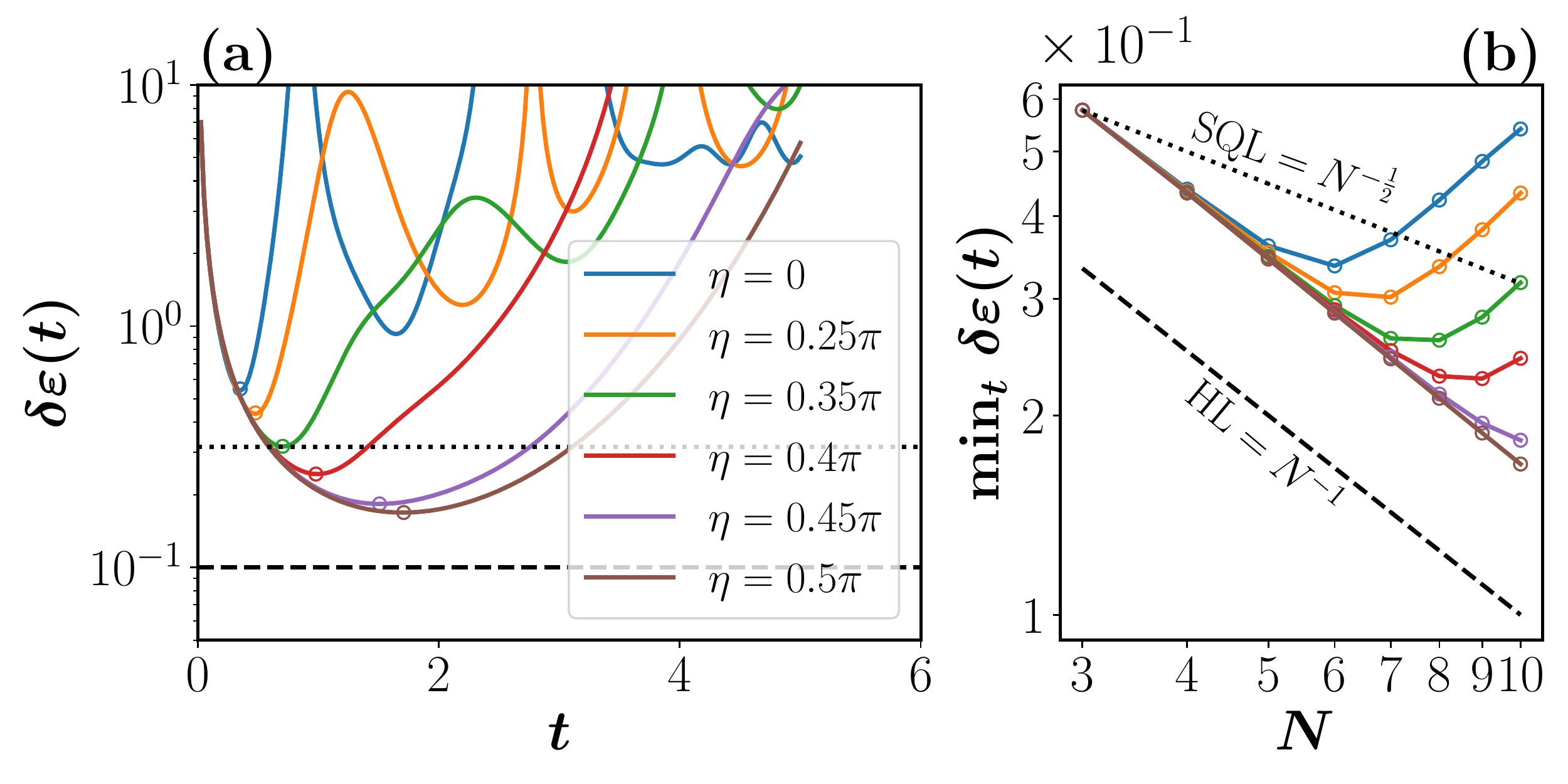}
		% Create a subtitle for the figure.
	\caption{The estimation error $\delta\varepsilon$ using a state prepared in our system of spin-1 particles constrained to a fixed volume $V=L^d$. Left: The error fails to reach even the the standard quantum limit (dotted black line, ${\rm SQL} = N^{-1/2}$) when $\eta \approx 0$, due to thermalisation of the spin-1 system. However, entanglement-enhanced estimation is possible when the model has QMBS ($\eta \approx \pi/2$) [plotted for $N=10$ spins]. Right: Due to the volume constraint, strong interactions between the spins typically inhibit the achievable error as the system size increases. However for perfect QMBS the error decreases uninhibited with the Heisenberg scaling $\sim N^{-1}$ (dashed black line). [Parameters: $L = 10$, $\omega = -\chi/N$, $\chi = 2$, $\lambda = 1$, $\gamma = 2$, $d=1$.]}
	\label{fig:min_error} 
	\end{center}
\end{figure}
We focus on our spin-1 model and -- to slightly simplify our scheme -- we choose $\omega = -\chi / N$ (since this gives cancellation between the terms proportional to $\hat{Q}^z = \frac{1}{2}\sum_{\vec{n}}\hat{S}_{\vec{n}}^z$ that appear in $\hat{H}_0$ and $\hat{H}_\text{nl} = \frac{\chi}{N} \hat{Q}^+ \hat{Q}^- = \frac{\chi}{N}[\hat{Q}^2 - (\hat{Q}^z)^2 + \hat{Q}^z]$). The scheme is as follows. (i) State preparation: starting from the spin coherent state $\ket{\theta = \frac{\pi}{2}, \phi = 0}$, evolve for a time $t$ by the total Hamiltonian, corresponding to the unitary $\hat{U}(\omega,\chi) = e^{-it[\hat{H}_0 (\omega) + \hat{H}_\text{int} + \hat{H}_\text{nl}(\chi)]}$. Note that this is a quantum chaotic Hamiltonian whose dynamics, in the absence of QMBS, is expected to lead to thermalisation; (ii) Parameter encoding: implement a unitary rotation via $\hat{U}_\varepsilon = e^{i\varepsilon \hat{Q}_y}$, where $\varepsilon \approx 0$ is the small parameter to be estimated; (iii) Echo measurement step: evolve for another time period $t$ by the total Hamiltonian, but with the signs of $\omega$ and $\chi$ reversed, i.e., by the unitary $\hat{U}(-\omega,-\chi)$. Note that the interaction term $\hat{H}_\text{int}$, which is responsible for potentially damaging interactions between the spins, is not reversed; (iv) Readout: finally, measure the observable $\hat{Q}_y$.

The final state after the evolution in steps (i)--(iii) is $\ket{\psi(2t)} = \hat{U}(-\omega,-\chi) e^{i \theta \hat{Q}_y} \hat{U}(\omega,\chi) \ket{\xi}$. For the observable measured in step (iv) the mean-squared error in the estimate of the parameter is: \begin{equation} \delta \varepsilon = \left| \frac{\sqrt{\text{Var}_{\ket{\psi(2t)}} \hat{Q}_y}}{\partial_\theta \langle \hat{Q}_y \rangle } \right|_{\varepsilon = 0} . \end{equation} Fig. \ref{fig:min_error}(a) shows this error as a function of the state preparation time $t$ for different values of the phase $\eta$ appearing in $\hat{H}_\text{int}$. When $\eta \approx 0$ the precision fails to reach even the standard quantum limit (dotted black line), due to thermalisation of the system during the state preparation and echo stages. However, we see significantly entanglement-enhanced estimation (i.e., beating the standard quantum limit) when the model has perfect QMBS at $\eta \approx \pm\pi/2$. Similarly, \ref{fig:min_error}(b) shows that -- due to the constraint that the spin system is confined to a fixed volume $V$ -- the precision is inhibited by interactions between the particles as the particle number $N$ (and hence the density $N/V$) is increased. On the other hand, the precision is enhanced with the Heisenberg scaling $\delta\varepsilon \sim 1/N$ when the model has perfect QMBS.\\

\textbf{Discussion-} 
In this work, we have presented a scheme to generate highly multipartite entangled states exploiting the presence of scarred subspaces in local Hamiltonians. Quantum many-body scars are currently the focus of a great attention, however potential applications are still mostly lacking. Here, we have proposed a metrological scheme which relies on the existence of the scars for entanglement enhanced phase-estimation. Most experiments showing QMBS so far have simulated the PXP model \cite{Ber-17, Su-22}, which does not straightforwardly fit into the SGA framework discussed here. However, evolution with the bare PXP Hamiltonian can generate multipartite entangled states at intermediate times \cite{Des-21} and using the prescription outlined in this work it can be stabilised and used for quantum enhanced metrology. We believe that the framework outlined in our work will inspire future works in quantum engineering of non-ergodic subspaces for applications in quantum technologies.
\\

\textit{Acknowledgements.---} 
JG is supported by a SFI-Royal Society University Research Fellowship and acknowledges funding from European Research Council Starting Grant ODYSSEY (Grant Agreement No. 758403). SP is supported by the Simons Foundation Grant No. 454943. SD acknowledges financial support from the SFI-EPRC joint project QuamNESS.
\newpage

\bibliographystyle{apsrev4-1}
\bibliography{references}

\appendix
\section{\label{app:deriving_nonlinear_H} Derivation of the non-linear Hamiltonian term}

As discussed in the main text, the SGA framework of QMBS consists of a subspace $\mathcal{S} \subset \mathcal{H}$ of the Hilbert space, and an operator $\hat{Q}^+$ that preserves the subspace and obeys the quasiparticle creation property \begin{equation} \hat{\mathcal{P}}_{\mathcal{S}} ( [\hat{H}, \hat{Q}^+] - \omega \hat{Q}^+ ) \hat{\mathcal{P}}_{\mathcal{S}} = 0 . \label{eq:Q_create} \end{equation}

Here we show that, within the general SGA framework, the non-linear entanglement generation term $\hat{H}_\text{nl} = \frac{\chi}{N}\hat{Q}^+\hat{Q}^-$ can emerge naturally in the scar subspace through a linear coupling to a highly detuned ancillary system \cite{Ben-13, Doo-16a}. 

To see this, consider a linear interaction with a bosonic mode of the form: \begin{equation} \hat{H}' = \hat{H} + \omega_a \hat{a}^\dagger \hat{a} + J (\hat{Q}^+ \hat{a} + \hat{Q}^- \hat{a}^\dagger) , \end{equation} where $\hat{a}^\dagger$, $\hat{a}$ are the bosonic creation and annihilation operators and $\omega_a$ is the frequency of the mode. Following Ref. \cite{Kli-00}, we consider the rotated Hamiltonian $e^{\hat{R}} \hat{H'} e^{-\hat{R}}$, where $\hat{R} = \frac{J}{\omega - \omega_a}(\hat{Q}^+ \hat{a} - \hat{Q}^- \hat{a}^\dagger)$. For a small rotation with $J \ll N |\omega - \omega_a|$ we can expand to second order in the small parameter: \begin{eqnarray} e^{\hat{R}} \hat{H'} e^{-\hat{R}} &\approx& \hat{H}' + [\hat{R}, \hat{H}'] + \frac{1}{2} [\hat{R}, [\hat{R}, \hat{H}'] ] + \hdots . \end{eqnarray} Restricting to the scar subspace and using Eq. \ref{eq:Q_create} then allows us to derive an effective Hamiltonian: \begin{eqnarray} \hat{\mathcal{P}}_{\mathcal{S}} ( e^{\hat{R}} \hat{H'} e^{-\hat{R}} ) \hat{\mathcal{P}}_{\mathcal{S}} &\approx& \hat{\mathcal{P}}_{\mathcal{S}} \bigg[ \hat{H} + \omega_a \hat{a}^\dagger \hat{a} + \frac{J^2}{\omega_a - \omega} \hat{Q}^+ \hat{Q}^- \nonumber\\ &&  \qquad  + \frac{J^2}{\omega_a - \omega} [\hat{Q}^+, \hat{Q}^-]\hat{a}^\dagger \hat{a}  \bigg] \hat{\mathcal{P}}_{\mathcal{S}} . \nonumber\end{eqnarray} If we also assume that the ancillary mode is in its vaccum state $\hat{a}^\dagger \hat{a} = 0$ then we finally have: \begin{equation} \hat{H}_\text{eff} = \hat{\mathcal{P}}_{\mathcal{S}} \left[ \hat{H} + \frac{J^2}{\omega_a - \omega} \hat{Q}^+ \hat{Q}^- \right] \hat{\mathcal{P}}_{\mathcal{S}} . \end{equation} in the scar subspace. Note that the strength of the non-linear term is $J^2/(\omega_a - \omega) \sim \mathcal{O}(1/N)$, due to the approximation condition $J \ll N |\omega - \omega_a|$. Defining $\chi = N J^2 / (\omega_a - \omega)$ makes this dependence on $N$ explicit, and gives the desired non-linear term $\hat{H}_\text{nl} = \frac{\chi}{N}\hat{Q}^+\hat{Q}^-$.

\end{document}